# THE NEXT-TO-LEADING ORDER OF THE DIFFERENTIAL CROSS-SECTION OF THE SUBPROCESS OF COMPTON SCATTERING OF QUARK-GLUON OF PROMPT PHOTON PRODUCTION IN PROTON-PROTON COLLISIONS AT NICA ENERGIES

**M.R.Alizada[1], A.I.Ahmadov[1,2]**

[1]Department of Theoretical Physics, Baku State University
Z.Khalilov 33, Az-1148 Baku, Azerbaijan, mohsunalizade@gmail.com
[2]Institute for Physical Problems, Baku State University,
Z.Khalilov 33, Az-1148 Baku, Azerbaijan

## ABSTRACT

In the presented article next-to-the leading order calculation of differential cross-section of the subprocess of Compton scattering $qg \to q\gamma$ of prompt photon production in proton-proton collisions at NICA energies has been carried out without and taking into account the longitudinal polarization of colliding protons.

Shown, that contribution of the next-to-leading order to differential cross-section is significant at the high energies of colliding protons and consist around 15% of leading order calculation. The influence of polarization of colliding protons on the next-to-leading order calculation is more significant than on the leading order calculation.

## I. INTRODUCTION

Photons are color-neutral and have a large mean free path even in dense hadronic matter, as well as in the environment of deconfined quarks and gluons. The production of prompt photons in proton-proton collisions is an important process for studying the dynamic structure of proton. Also, the process of prompt photon production plays an important role in determining the gluon distribution in the proton and testing some aspects of perturbative quantum chromodynamics (pQCD). Especially, the investigation of the subprocesses involving sea quarks allows to study of the structure of the proton in more detail [1-3].

At early, theoretical and experimental investigations of the prompt photon production in the proton-proton collisions in leading order (LO) and next-to leading order (NLO) was been carried out at the LHC and American Tevatron energies [4,5]. It is shown that the subprocesses of Compton scattering of a quark-gluon $qg \to q\gamma$ and the annihilation of a quark-antiquark pair $q\bar{q} \to g\gamma$ are the basic mechanisms for the production of prompt photons. The mixed QCD Compton scattering subprocess is determined to be the dominant process of prompt photon production in proton-proton collisions.

The differential cross-section of subprocess Compton scattering of quark-gluon $qg \to q\gamma$ can be calculated in pQCD, using the factorization theorem, which separates the short-range particle interactions from the long-range hadronic effects [6-8].

In the [10] are presented results of theoretical studies of the isolated production of prompt photons at the DESY HERA collider in the NLO quantum chromodynamics. The authors used analytical methods to calculate the differential cross-sections of various subprocesses and applied isolating cross-section, corresponding to the experimental conditions. It was shown that the differential cross-section is sensitive to the parton distribution function (PDF) of both the photon and the quark, but the gluon distribution is difficult to measure.

In the article [11] is described the Monte Carlo method for calculating the production of prompt photons beyond the leading logarithm approximation. This method combines analytical and numerical integration methods for handling the higher-order subprocess, as well as collinear and soft singularities. The method is applied to various observables and compared with experimental data, showing good agreement and reduced scale dependence.

In the articles [12,13] are presented the simulation of prompt photon production at proton-proton collisions at LHC and American Tevatron energies was performed in using the program JETPHOX, based on the Monte Carlo method.

The accuracy of theoretical predictions depends on the order of the perturbative expansion, which is usually expressed in terms of the strong coupling constant $\alpha_s$. LO approximation, which corresponds to the lowest power of $\alpha_s$, is often insufficient to describe the experimental data, especially at large transverse momenta of photons $p_T$. Therefore, it is necessary to use higher-order corrections, such as NLO, which take into account the effects of one-loop virtual diagrams and real production of additional particles.

However, NLO calculations are more complicated than LO calculations, because they include ultraviolet and infrared divergences, which need to be regularized and renormalized. NLO results depend on the choice of renormalization and factorization scales, which introduce theoretical uncertainties, which need to be estimated [9].

We investigated differential cross-section of prompt photon production in proton-proton collision at NICA energies. Research into the production of prompt photons in proton-proton collisions at NICA energies has its advantages. The energy of NICA (10-27 GeV) does not allow the production of many elementary particles, which makes it difficult to accurately determine the differential cross-section of prompt photons production. From this point of view, the studies that are planned to be carried out at the NICA experimental facility are important [14-16].

Early, we studied the process of prompt photon production in proton-proton collisions at NICA energies in LO approximation without and taking into account the longitudinal polarization of colliding protons [17-19]. We considered such subprocesses as mixed Compton scattering of quark-gluon $qg \to q\gamma$, annihilation of quark-antiquark pair $q\bar{q} \to g\gamma$ and bremsstrahlung $qq \to qq\gamma$. We found out that the differential cross-sections of the process of prompt photon production in proton-proton collisions at center-of-mass energies of 10 GeV are determined by approximately equal contributions of the subprocesses of mixed chromo-electrodynamics Compton scattering of quark-gluon $qg \to q\gamma$ (more 50%) and annihilation of quark-antiquark pair $q\bar{q} \to g\gamma$ (more 47%), while the contribution of bremsstrahlung $qq \to qq\gamma$ can be considered negligible (0.03%). These values are lower than those found for the LHC and American Tevatron energies. Moreover, despite the low collision energies, Compton scattering of quark-gluon $qg \to q\gamma$ remains dominant. We also have shown that the prompt photons produced in pure electrodynamics Compton scattering constitute 10% of the prompt photons produced in mixed Compton scattering [19].

The presented article is devoted to the calculations of differential cross-sections for Compton scattering, performed in the NLO without and taking into account the longitudinal polarization of colliding protons, and compares with calculations in LO.

## II. NEXT-TO-LEADING ORDER DIFFERENTIAL CROSS-SECTION OF THE SUBPROCESS COMPTON SCATTERING OF QUARK-GLUON $qg \to q\gamma$

We used FeynCalc for generation of Feynman diagrams of the subprocess of mixed quark-gluon Compton scattering $qg \to q\gamma$. In this case 773 one-loop diagrams were generated. Among the diagrams, the ones with heavy particles (H, G, Z, W, t, c, b, μ, τ, e), which are not registered at NICA SPD, the ones with triangular loops of fermions and vector bosons, which vanish by the Furry theorem [20], the ones with "seagulls" and the ones with loops on external legs, which are excluded in the renormalization scheme on the mass shell, as well as the ones with photon-gluon transitions, which do not conserve color, were not taken into account. As a result, we obtained the following 11 diagrams, which contribute to the prompt photons production in NLO at NICA energies. Feynman diagrams of the process are shown in Figure 1 in the Appendix.

The square of the modulus of the matrix element for the one-loop correction subprocess $qg \to q\gamma$, with applied dimensional regularization and averaged over the spin of the initial particles, has the form:

$$|\bar{M}|^2 = \frac{0.0012\alpha_e \alpha_s^3 e_q^2 (\hat{s}^3 + \hat{s}^2\hat{u} + \hat{s}\hat{u}^2 + \hat{u}^3)}{\hat{u}^2(\hat{s}+\hat{u})},$$

where $\alpha_s$ - is the strong coupling constant at the factorization scale, which we choose equal to $\hat{s}$, $\alpha_e$ - is the fine-structure constant, $e_q$ - is the electric charge of the quark.

The differential cross-section of the subprocess in parton level was defined as in [21].

$$\frac{d\hat{\sigma}_{NLO}(qg\to q\gamma)}{d\hat{t}} = \frac{1}{16\pi\hat{s}^2}|\bar{M}|^2 \quad (1)$$

To calculate the differential cross-section at the hadron level, we used the parton distribution function (PDF) in NLO level $G_{q_1/h_1}(x_1)$ and $G_{q_2/h_2}(x_2)$ in accordance with [22] at the factorization scale $\hat{s}$:

$$d\sigma = \int d\hat{\sigma} \cdot G_{q_1/h_1}(x_1)G_{q_2/h_2}(x_2)dx_1dx_2.$$

To ensure consistency in the evaluation of parton distribution functions (PDFs) under scale variation, we used LHAPDF version 6.5.5 to access modern PDF sets with built-in support for scale-dependent evolution [23]. This allows us to study how the variation in scale impacts both the normalization and shape of differential distributions.

In particular, for the differential distributions by rapidity $y$ and square of transverse momentum $p_T$, we obtain the following expressions:

$$\frac{d\sigma}{dy} = \int (-t)G_{q_1/h_1}(x_1)G_{q_2/h_2}(x_2)dx_1dx_2\frac{d\hat{\sigma}}{d\hat{t}}$$
$$\frac{d\sigma}{dp_T^2} = \int (\frac{-t}{2p_T^2})G_{q_1/h_1}(x_1)G_{q_2/h_2}(x_2)dx_1dx_2\frac{d\hat{\sigma}}{d\hat{t}} \quad (2)$$

Differential cross-section of prompt photon production in collision of longitudinally polarized protons was been calculated. For this longitudinal polarization of colliding quarks is taken into account as in [24]:

$$\mathbf{U(p_1)\bar{U}(p_1) = \frac{1}{2}(1-\lambda_1\gamma_5)(\hat{p}_1+m_1)},$$

where $\lambda_1$, $m_1$ and $p_1$ of helicity, mass and transverse momentum of quark, correspondingly.

To take into account the polarization of the initial gluon, the four-dimensional vectors of the initial particles are written as in [25]:

$$\hat{e}^{\lambda_2}(k_1) = \frac{(1+\lambda_2\gamma^5)\hat{k}_2\hat{q}\hat{k}_1+(1-\lambda_2\gamma^5)\hat{k}_1\hat{q}\hat{k}_2}{4\sqrt{(k_1k_2)(k_1q)(k_2q)}},$$

where $\lambda_2$ helicity of initial gluon.

To calculation of the differential cross-section of processes with polarized colliding protons a polarized PDF was used [26]:

$$\Delta u_v(x, Q_0^2) = x^{-0.441}(1-x)^{3.96}(0.928 + 0.149x^{0.5} - 1.141x + 11.612x^{1.5})$$

$$\Delta d_v(x, Q_0^2) = x^{-0.665}(1-x)^{4.46}(-0.038 - 0.43x^{0.5} - 5.260x + 8.443x^{1.5})$$

$$\Delta G(x, Q_0^2) = x^{-1.17}(1-x)^{5.33}(0.03 - 1.71x^{0.5} + 3.01x + 43.5x^{1.5})$$

It is known that the fraction of photon-type partons in a proton can be comparable to the fraction of sea quarks at large values $x$ [26,27]. The distribution of polarized photons in the proton was taken into account as in [26]:

$$G_{\gamma/p}(x, Q_0) = \frac{\alpha_e}{2\pi}\left(A_u e_u^2 \tilde{P}_{\gamma q} \otimes u^0(x) + A_d e_d^2 \tilde{P}_{\gamma q} \otimes d^0(x)\right),$$

where: $A_i = \ln\left(Q_0^2/m_i^2\right)$, and $\tilde{P}_{\gamma q} = \frac{1+(1-z)^2}{z}$ splitting function, $Q_0$ factorization scale.

The matrix elements and Mandelstam invariants will be identical, without the longitudinally polarization case. Taking into account the longitudinal polarization of the initial particles, for the square of the matrix element averaged over the spin of the initial particles, for the subprocess $qg \to q\gamma$ of one-loop correction we get:

$$|\bar{M}|^2 = \frac{0.0012\alpha_e\alpha_s^3 e_q^2}{\hat{s}\hat{u}^2(\hat{s}+\hat{u})}\{\hat{s}^4(1-2\lambda_1\lambda_2+\lambda_2^2) + \hat{s}^3\hat{u}(1-2\lambda_1\lambda_2+\lambda_2^2) + $$
$$+\hat{s}^2\hat{u}^2(1+2\lambda_1\lambda_2+\lambda_2^2) + \hat{s}\hat{u}^3(1+2\lambda_1\lambda_2+\lambda_2^2)\}$$

The differential cross-section of the subprocess in parton and hadron level was defined as (1) and as (2), correspondingly.

Doublespin asymmetry of subprocesses $qg \to q\gamma$ has been calculated as in [28]:

$$A_{LL} = \frac{\sigma^{\uparrow\uparrow} - \sigma^{\uparrow\downarrow}}{\sigma^{\uparrow\uparrow} + \sigma^{\uparrow\downarrow}}$$

where $\sigma^{\uparrow\uparrow}$ и $\sigma^{\uparrow\downarrow}$ - cross sections of processes calculated with, correspondingly, directed and oppositely directed polarizations of colliding protons.

To compare differential cross-sections at the parton level, calculated in the LO and NLO, we used the ratio $\hat{R} = \frac{d\hat{\sigma}_{NLO}}{d\hat{\sigma}_{LO}}$ and find the expression $\hat{R} \approx 0.005\alpha_s$. The hadron level relationship $R = \frac{d\sigma_{NLO}}{d\sigma_{LO}}$ is determined using PDF.

The influence of the polarization of colliding protons on the differential cross-section of the process of prompt photons production was studied by determining the $= \frac{d\sigma_{NLO,POL}}{d\sigma_{NLO}}$.

## III. NUMERICAL RESULTS AND THEIR DISCUSSION

We investigate the dependences of differential cross-section of prompt photon production on kinematic parameters: energy of colliding protons $\sqrt{s}$, transverse momentum $p_T$, cosine of scattering angle of photon, rapidity $y$ and $x_T$.

Figure 2 shows the dependences of the total cross-sections of the subprocess $qg \to q\gamma$ calculated in the LO and the NLO on the sum of energies of colliding protons $\sqrt{s}$, with the following restrictions on the transverse momentum of the photon $p_T$ and its rapidity: $p_T \geq 0.5$ and $-2 \leq y \leq 2$.

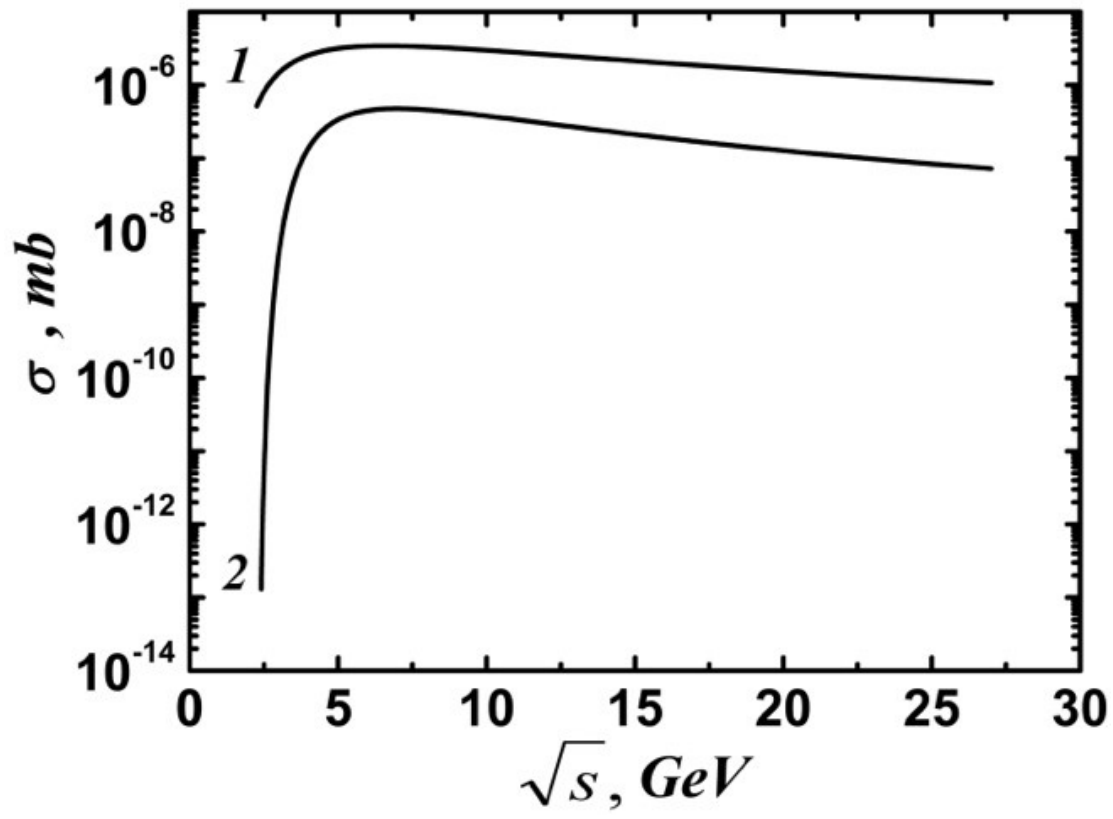

Fig.2 The dependence of the total cross-section of the subprocess $qg \to q\gamma$ calculated in the LO (curve 1) and the NLO (curve 2) on the sum of the energies of the colliding protons $\sqrt{s}$

As can be seen from Figure 2, character of dependence of the total cross-sections of the subprocess $qg \to q\gamma$ on the sum of energies of colliding protons $\sqrt{s}$ calculated in the LO and the NLO is same. The total cross-section increases with an increase in the sum of energies of the colliding protons and reaches a maximum at the $\sqrt{s}$=4.6 GeV. With a further increase in energy, the total cross section begins to decrease. It can be assumed, with an increase in energy, the according to the Lorentz transformation the spherical shape of the particles changes to a disk shape. In this case, the particles from the dense state of their components pass to a less dense state. In this form, when protons collide, the collision of their components is less probable and, therefore, the differential collision cross-section decreases. The energy of 4.6 GeV can characterize the energy of strong interaction, as well as the minimum energy required for the transition of particle from spherical form to the disk shape. Another assumption is that with an increase in the speed (energy) of colliding particles, the interaction time of the particles (the strong coupling constant $\alpha_s$) decreases, and therefore the differential collision cross section decreases. This is consistent with the prediction of pQCD.

At small values of energy of colliding protons $\sqrt{s}$. the total cross-section calculated in the NLO is significantly smaller than the total cross-section calculated in the LO, but becomes approximately comparable at high values of $\sqrt{s}$. This indicates that the NLO corrections are important for the accuracy of the calculation, especially at high values of energy of colliding protons $\sqrt{s}$.

Figure 3 shows the dependence of the differential cross-section $\frac{d\sigma}{dp_T^2}$ calculated in the LO and the NLO on the transverse momentum $p_T$ of the produced prompt photons. As seen from Figure 3 with increasing transverse momentum of the photon $p_T$, the differential cross-section decreases rapidly.

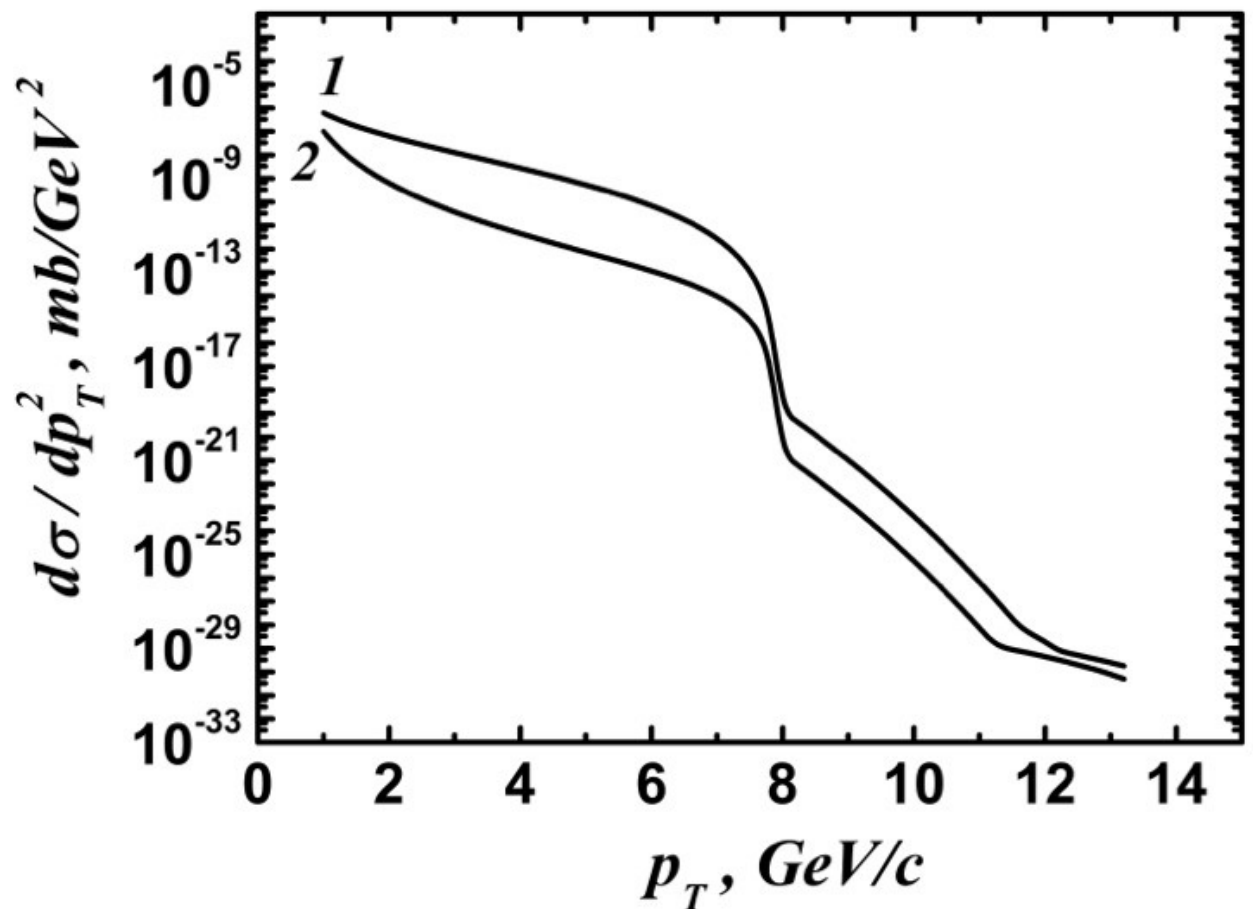

Fig.3 The dependence of the differential cross-section $\frac{d\sigma}{dp_T^2}$ of the subprocess $qg \rightarrow q\gamma$ calculated in the LO (curve 1) and the NLO (curve 2) on the transverse momentum $p_T$ of the prompt photons at $\sqrt{s}$=27 GeV.

This is because photon emission is more likely to occur at small angles with respect to the colliding quark, implying a smaller $p_T$. As the transverse momentum of photons increases, photon production becomes increasingly suppressed by the strong coupling constant $\alpha_s$ and the PDF. This behavior is consistent with the predictions of pQCD, which describes the interaction between quarks and gluons at high energies. In pQCD, the probability of the hard scattering process decreases as the transfer momentum (in this case, the photon transverse momentum $p_T$) increases. The rapid decrease may also be due to limitations of the pQCD approach at very high $p_T$. At these energies, nonperturbative effects become significant and pQCD calculations may be inaccurate. The specific shape of the curve and the rate of decrease may provide insight into the underlying mechanisms of this subprocess, such as the PDFs of the colliding particles or the specific contributions of different Feynman diagrams.

The dependence of the differential cross-section on the transverse momentum $p_T$ of the produced prompt photons. shows that at the considered interval of variation $p_T$ the differential cross-section calculated in NLO is smaller than LO. At small and large values of $p_T$, the difference between the differential cross-sections calculated in LO and NLO is small.

Figure 4 shows the dependence of the differential cross-section on the cosine of the scattering angle of prompt photons.

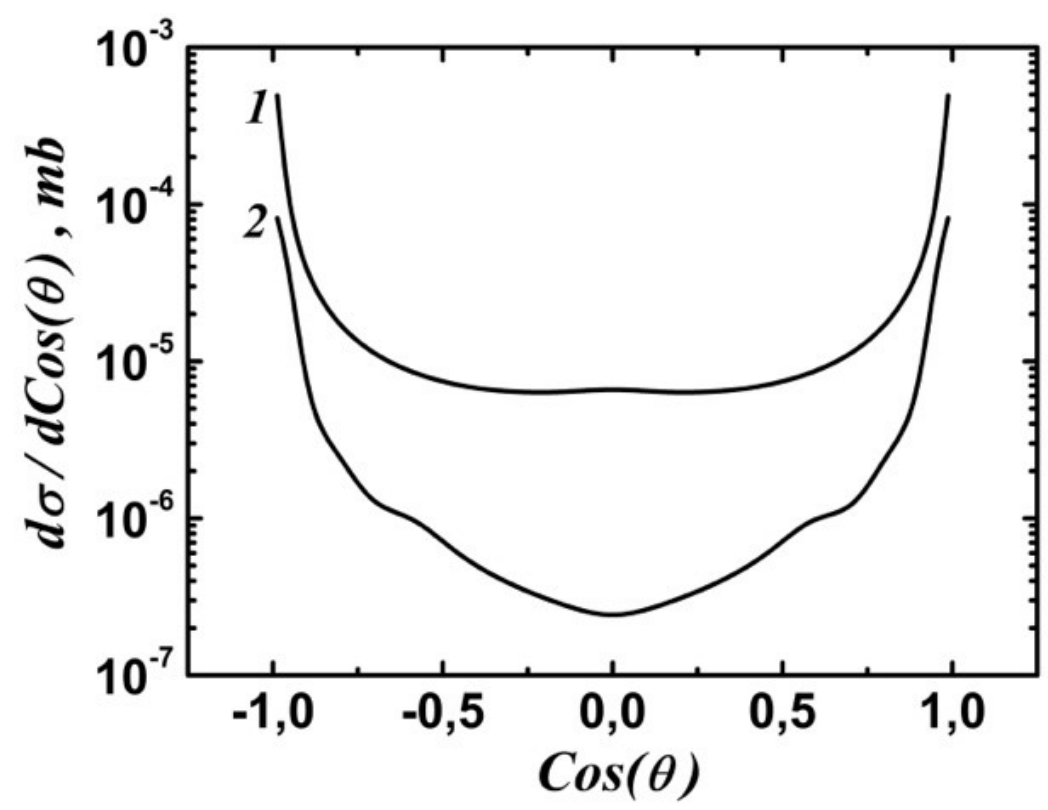

Fig.4 The dependence of the differential cross-section $\frac{d\sigma}{dCos(\theta)}$ of the subprocess $qg \to q\gamma$ calculated in the LO (curve 1) and the NLO (curve 2) on the cosine of the scattering angle of prompt photons at $\sqrt{s}$=27 GeV.

The dependence of the differential cross-section on the cosine of the scattering angle of the prompt photon for the subprocess $qg \to q\gamma$ calculated in the LO and the NLO have a similar character. The differential cross-section has the largest value when the angle of scattering of photon approaches 16 or 164 degrees. This means that the photon is most likely scattered in the direction of the axe of collision of protons. The differential cross-section becomes smaller with the increase of the scattering angle of photon.

Figure 5 shows the dependence of the differential cross-section $\frac{d\sigma}{dy}$ of the subprocess $qg \to q\gamma$ calculated in the LO and the NLO on the rapidity $y$ of the prompt photons.

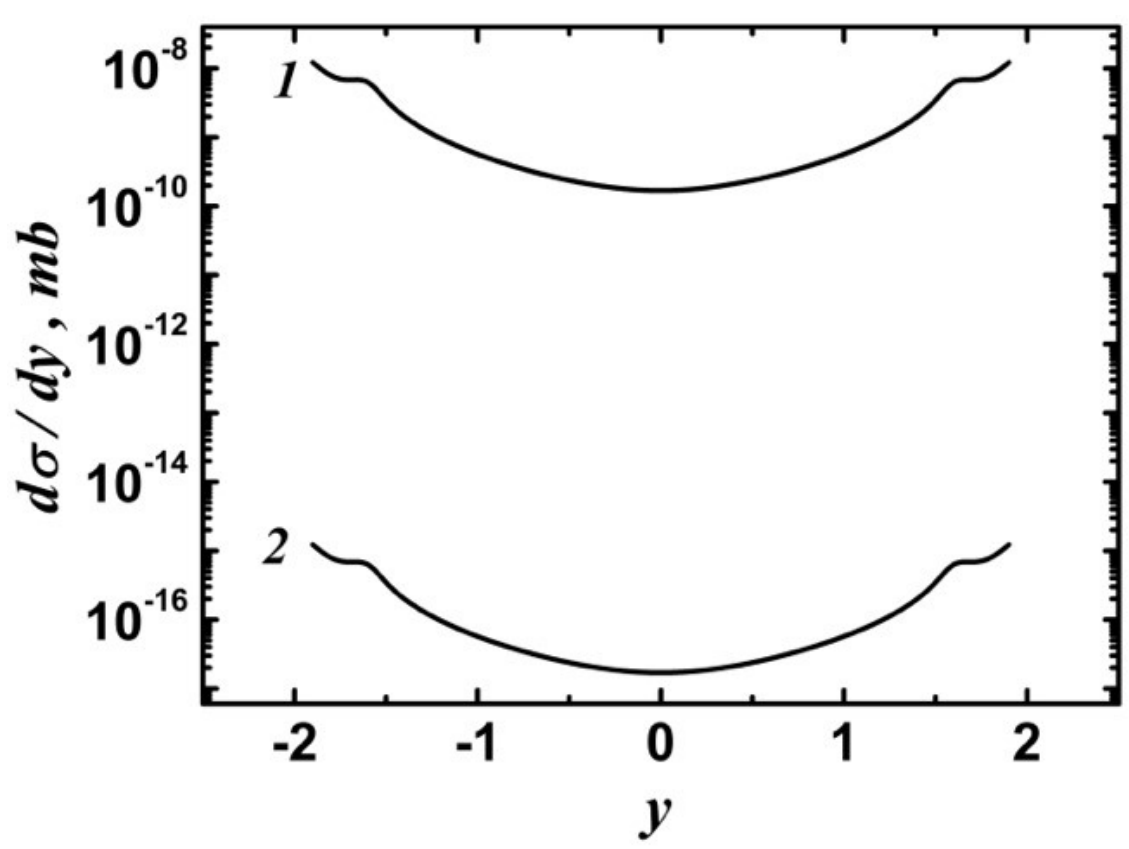

Figure 5 The dependence of the differential cross-section $\frac{d\sigma}{dy}$ of the subprocess calculated in the LO (curve 1) and the NLO (curve 2) on the rapidity $y$ of the prompt photons at $\sqrt{s}$=27 GeV.

It can be seen from Figure 5 that the differential cross-section reaches the maximum value at $y$=±1.95 and decreases with the change of $y$. The minimum value of the differential cross-section is observed at $y$=0.

This behavior suggests that the process is more probable at higher speed, when the colliding particles have a larger longitudinal momentum. Since the differential cross section has a symmetric shape around $y$=0, this means that the subprocess is invariant under the transformation $y$→-$y$. This is expected from the parity and charge conjugation symmetries of QCD and QED. Reaching a minimum value at $y$=0, which suggests that the process is less probable when the quark and gluon have the same longitudinal momenta, this means that the subprocess is more likely to produce a quark and a photon with equal and opposite rapidities, or, equivalently, with the same transverse momentum and zero longitudinal momentum, is less likely. This is consistent with the law of conservation of momentum in collisions.

Figure 6 shows the dependence of the differential cross-section of the subprocess $qg \rightarrow q\gamma$ calculated in the LO and the NLO on $x_T$.

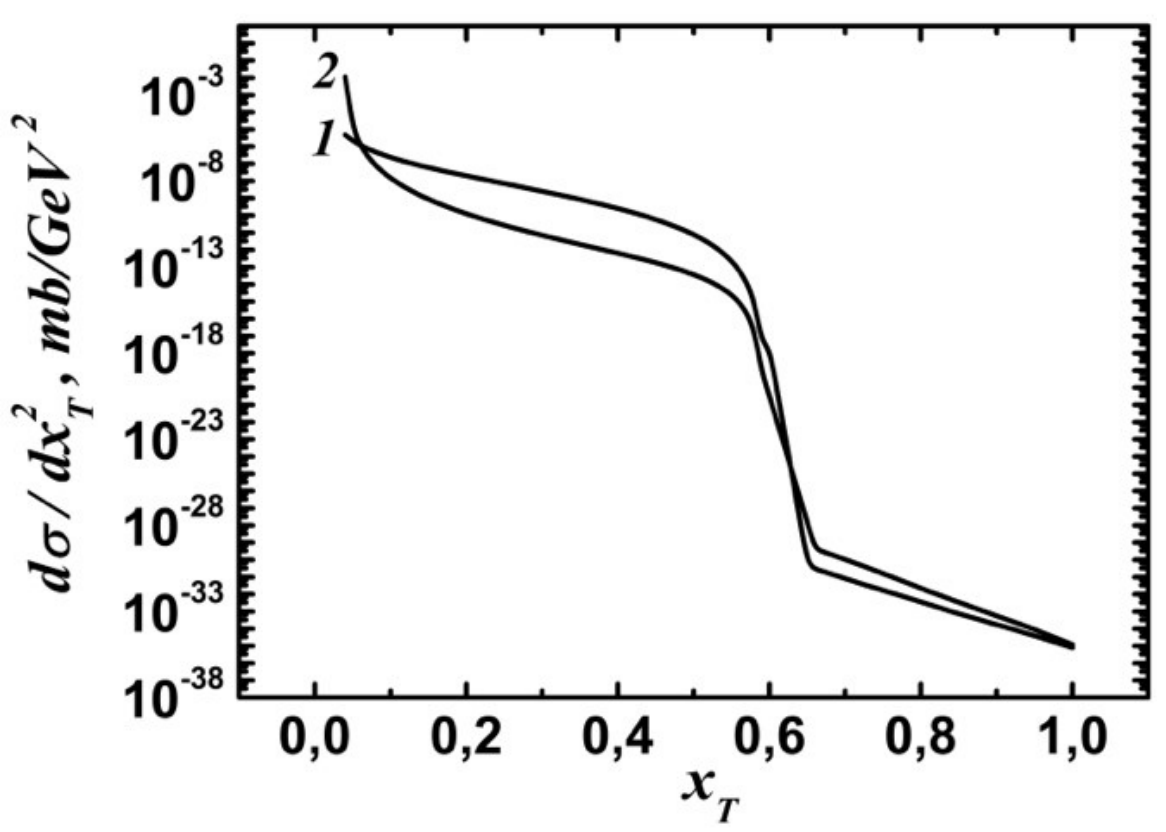

Fig.6 The dependence of the differential cross-section $\frac{d\sigma}{dy}$ of the subprocess $qg \rightarrow q\gamma$ calculated in the LO (curve 1) and the NLO (curve 2) on $x_T$ at $\sqrt{s}$=27 GeV.

The dependence of the differential cross-section on $x_T$ has a similar character as the dependence of the differential cross-section on the transverse momentum $p_T$ of produced prompt photon. The trend is similar to the $p_T$ dependence since $x_T$ is directly proportional to $p_T$ ($x_T = 2p_T/\sqrt{s}$).

Thus, as expected all dependences of the differential cross-section calculated in the LO, has a larger value than all dependences of the differential cross-section calculated in the NLO. This is expected in pQCD, where the higher-order terms introduce more significant radiative corrections at higher energies of the partons.

Consequently, NLO calculation is significant value at high energies of colliding protons.

Investigation of the dependences of ratio $R = \frac{d\sigma_{NLO}}{d\sigma_{LO}}$ on the sum of the energies of the colliding initial particles $\sqrt{s}$, the transverse momentum $p_T$ of the prompt photons, the cosine of the scattering angle $Cos(\theta)$, the rapidity $y$ of the prompt photons show that the NLO corrections play an increasingly important role in describing the process of prompt photon production.

Figure 7(*a,b,c,d*) shows the dependences of the ratio $R = \frac{d\sigma_{NLO}}{d\sigma_{LO}}$ on the sum of the energies of the colliding particles $\sqrt{s}$ (*a*), the transverse momentum $p_T$ of the prompt photons (*b*), the rapidity $y$ (*c*) and $x_T$ (*d*) at $\sqrt{s}$=27 GeV.

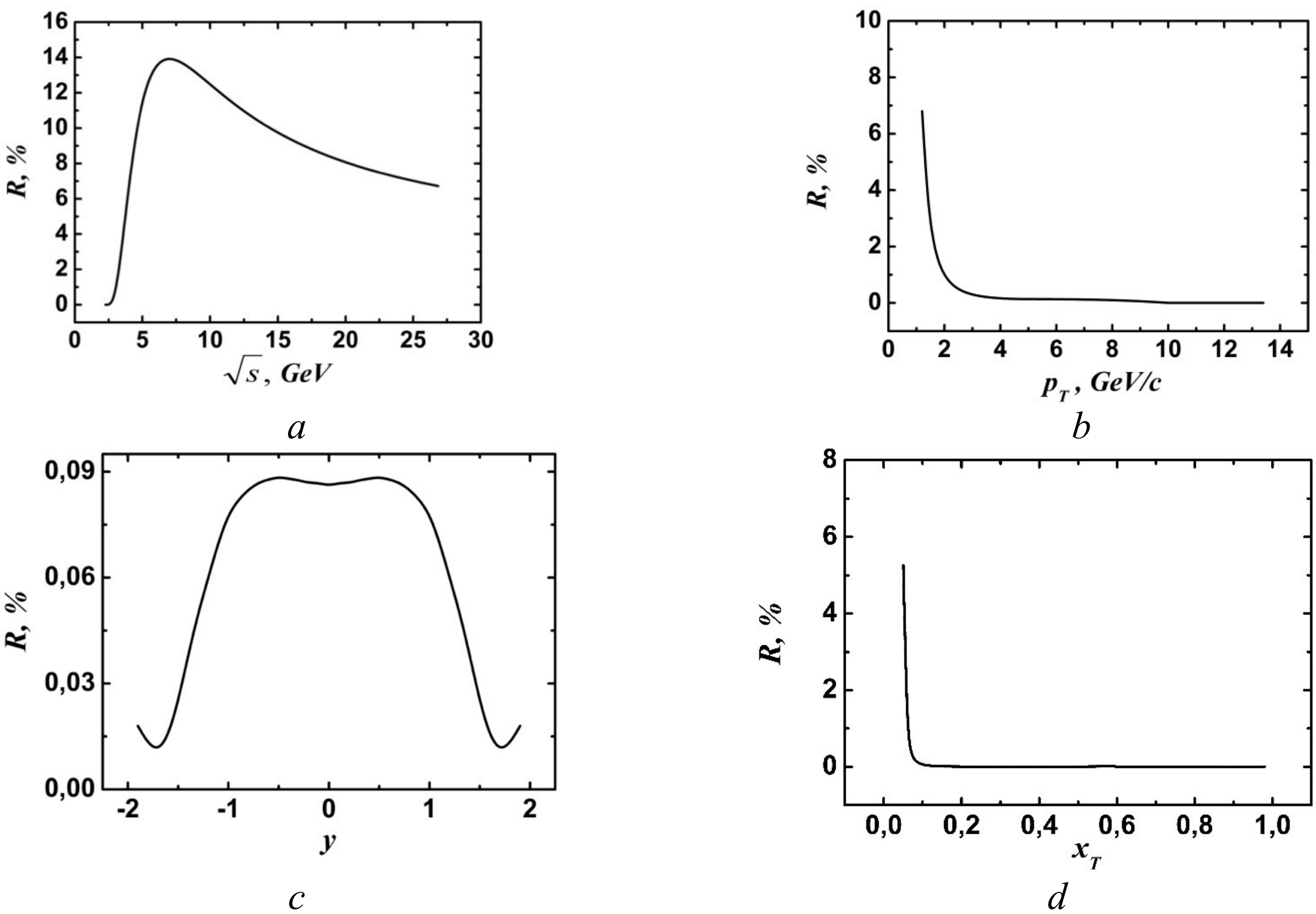

Fig.7(*a,b,c,d*) The dependencies of the ratio $R = \frac{d\sigma_{NLO}}{d\sigma_{LO}}$ on the sum of the energies of the colliding particles $\sqrt{s}$ (*a*), the transverse momentum $p_T$ of the prompt photons (*b*), the rapidity $y$ (*c*) and $x_T$ (*d*) at $\sqrt{s}$=27 GeV.

As seen from Figure 7(a) that $R$ increases with the increase of the sum of energies of the collision of particles $\sqrt{s}$. $R$ reaches the maximum when $\sqrt{s}$=6.5 GeV, and then decreases with the increase of $\sqrt{s}$. As can be seen from Figure 7(*a*), the contribution of the NLO is significant at high energies of the colliding protons. This is consistent with Figure 2.

With the increase of the transverse momentum $p_T$, the value of $R$ decreases in the interval $p_T \epsilon[1;13\ 4.5]$. $R$ has the maximum value equal to 6.5% at $p_T$=1 GeV/c (Figure 7(*b*)). This behavior is expected because higher-order corrections are generally more significant at lower $p_T$, where soft gluon emissions and virtual corrections play a dominant role. At high $p_T$, the perturbative expansion becomes more stable, reducing the relative contribution of NLO corrections.

From the dependence of the ratio $R$ on the rapidity $y$ photon (Figure 7(*c*)), it can be seen that $R$ monotonically decreases with the increase of the absolute values of $y$. The means that the corrections of the NLO become more important in the direction of the central region ($y$=0) and less significant in the forward and backward directions (high absolute values of $y$). This suggests that NLO corrections are more pronounced in the central rapidity region where phase space for gluon radiation and virtual corrections is larger. - At large rapidities ($|y| \sim 2$), the suppression of NLO effects could be due to the dominance of leading-order processes in the forward/backward region.

Dependence on variable $x_T = 2p_T/\sqrt{s}$ is $R$ decreases rapidly with $x_T$. This follows the same reasoning as for $p_T$, where NLO corrections are more significant at small $x_T$, but their impact diminishes as $x_T$ increases, meaning the perturbative expansion stabilizes.

In the Figure 8(*a,b,c,d,e*) are present the dependences of differential cross-section in the hadron level of by taking in to account longitudinal polarization of colliding protons on sum energy of colliding protons $\sqrt{s}$, transverse momentum $p_T$, cosine of scattering angle, rapidity of prompt photon $y$ and $x_T$ at $\sqrt{s}$=27 GeV.

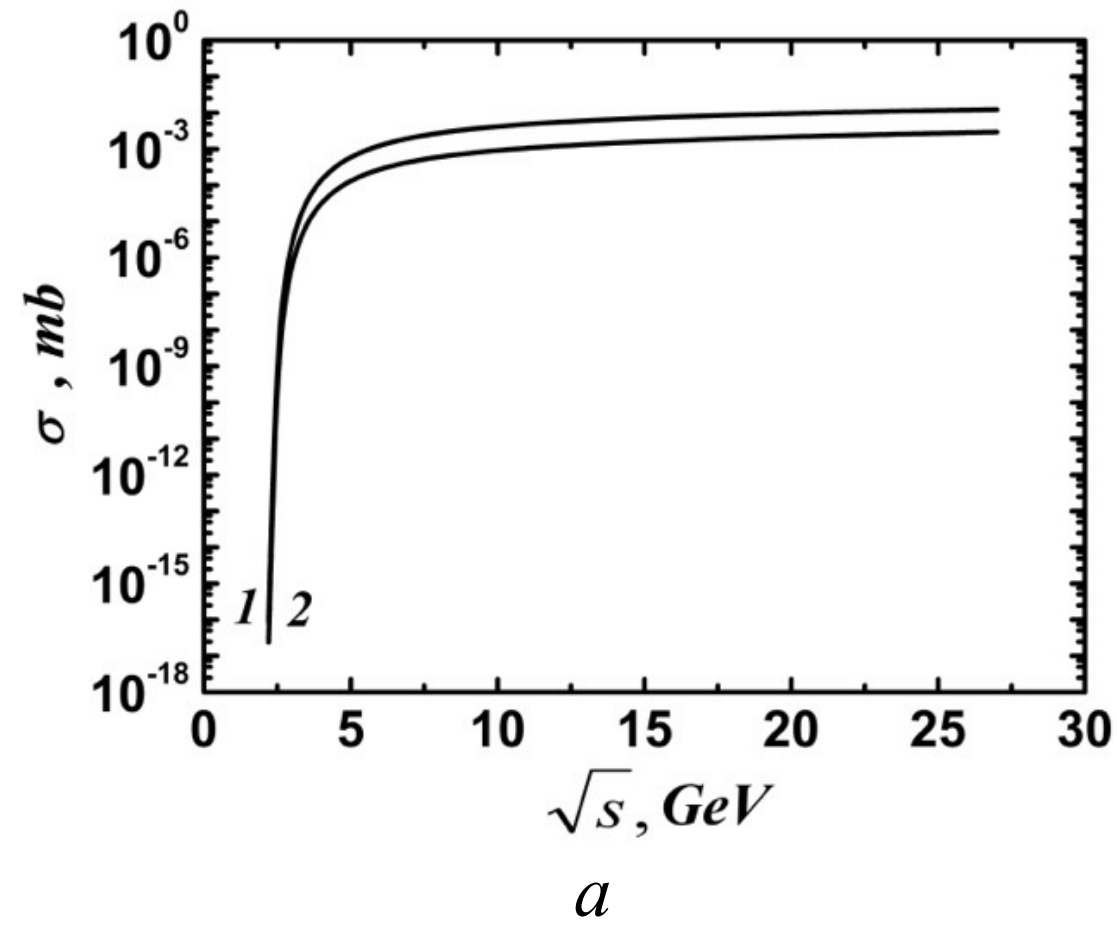

*a*

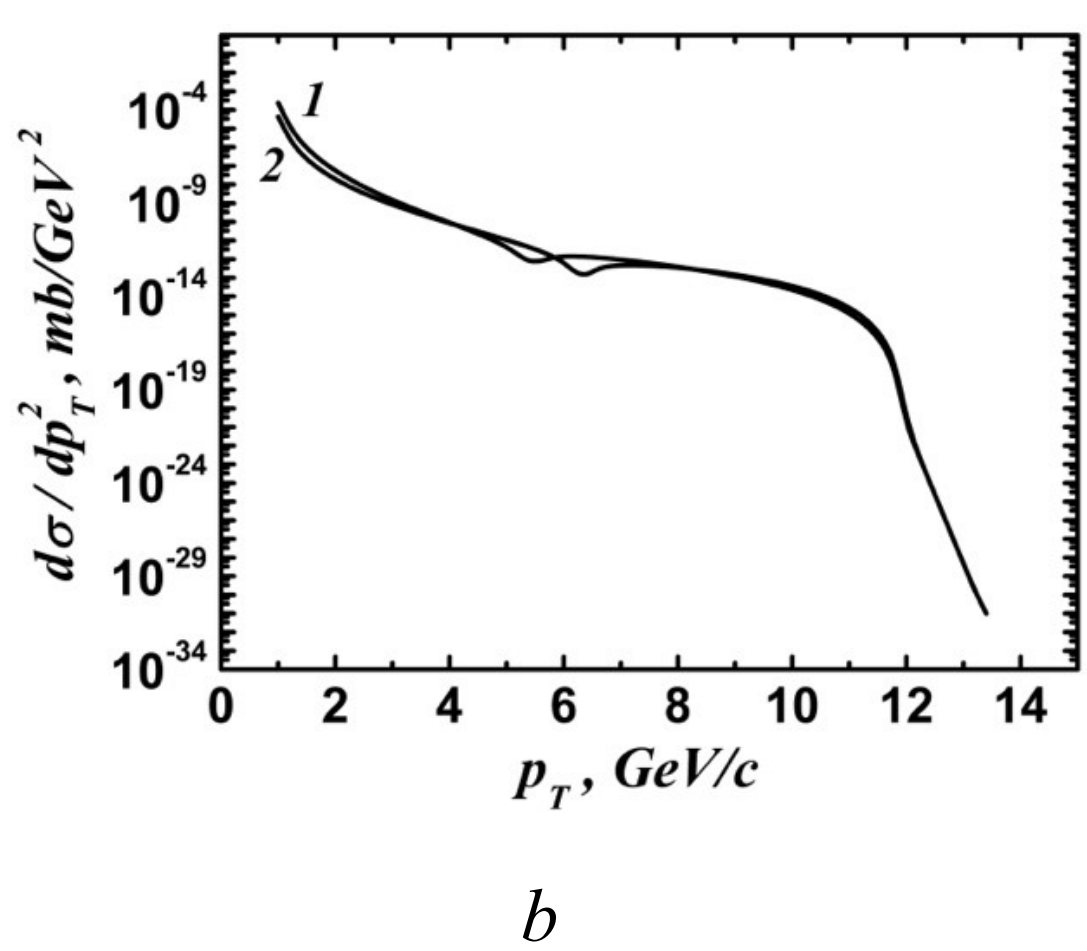

*b*

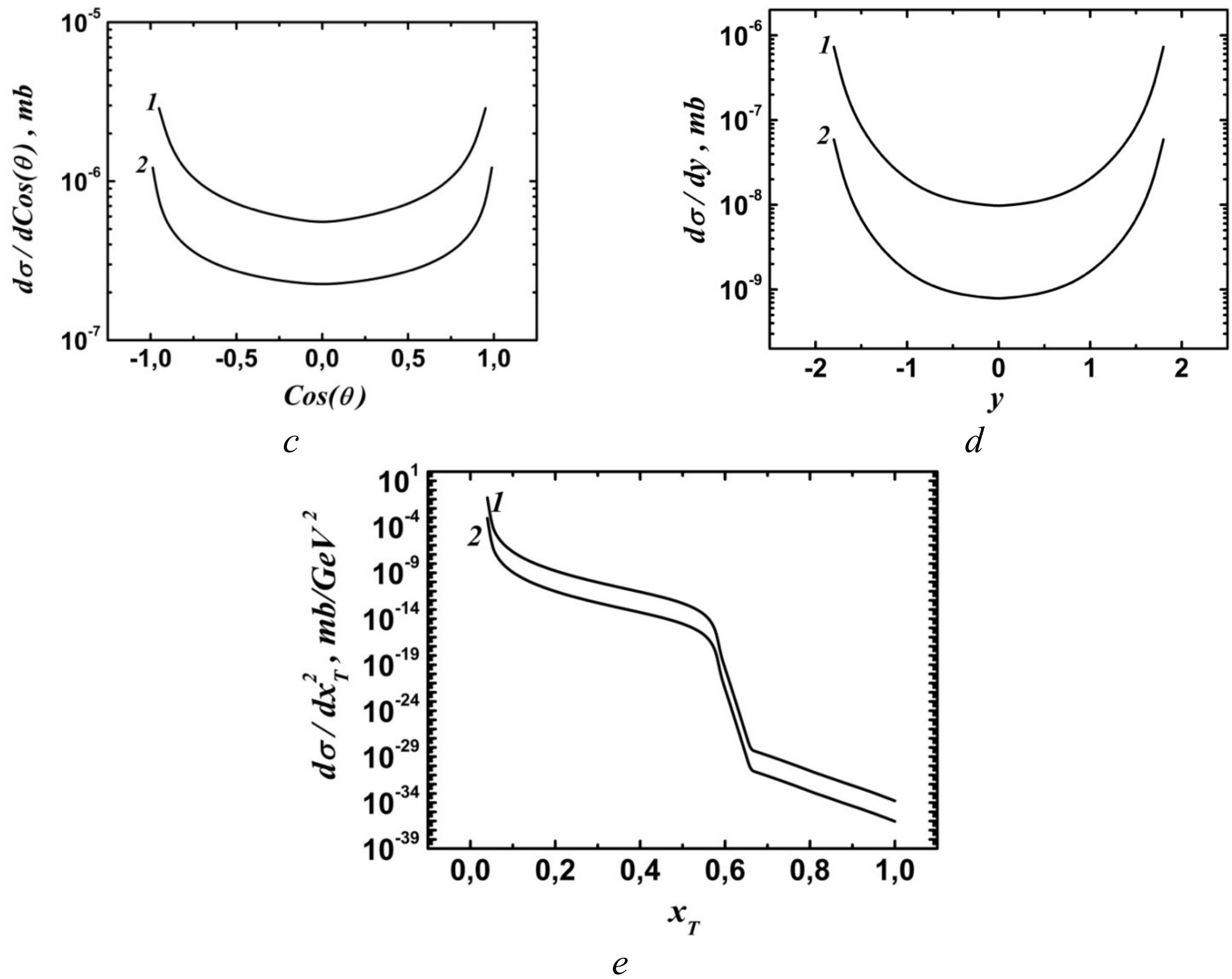

*c* *d*

*e*

Fig.8(*a,b,c,d,e*) The dependence of differential cross-section subprocess $qg \to q\gamma$ calculated in NLO with taking account longitudinal polarization of colliding protons on sum energy of colliding protons, $\sqrt{s}$ (*a*), transverse momentum $p_T$ (*b*), cosine of scattering angle (*c*), rapidity of prompt photon $y$ (*d*) and $x_T$ (*e*) at $\sqrt{s}$=27 GeV at polarization order $P_1$ ,$P_2$=0.9, -0.9 (curve 1) and at $P_1$=$P_2$=±0.9 (curve 2).

As can be seen from Figure 8(*a,b,c,d,e*), polarization did not change the character of the dependences of differential cross-section on sum energy of colliding protons, $\sqrt{s}$, transverse momentum $p_T$, cosine of scattering angle, rapidity of prompt photon $y$ and $x_T$. In the opposite polarized case, the differential cross-section increases, and in the directional polarized case, the differential cross section decreases, which is similar to LO calculation.

As shown in Figure 8(*a*), an increase in the sum energy of colliding particles $\sqrt{s}$ in the interval [2;27] GeV leads to an increase in the total cross-section of the subprocess $qg \to q\gamma$.

A study of the dependence of the differential cross section on the transverse momentum $p_T$ of the produced prompt photons (Figure 8(*b*)) shows a rapid decrease as $p_T$ increases.

From Figure 8($c$) it can be seen that the dependence of the differential cross-section on the cosine of the scattering angle of prompt photons is most pronounced when the scattering angle approaches approximately 16 or 164 degrees, indicating a higher probability of photon production along the collision axis of protons. Conversely, as the scattering angle changes between 16 and 164 degrees, the differential cross section decreases.

The maximum differential cross-section appears at $y=\pm1.95$ (Figure 8($d$)), and decreases as y is changed in the interval (1.95, -1.95), reaching a minimum at $y=0$.

Moreover, the trend of the differential cross section versus $x_T$ is very similar to the trend of the transverse momentum $p_T$ (Figure 8($b$), Figure 8($e$)).

Figures 9($a,b,c$) show the dependences of the doublespin asymmetry of Compton scattering of quark-gluon in NLO on the transverse momentum of photons $p_T$, $x_T$ and a 3D plot of the dependence of the doublespin asymmetry on the polarization orders $P_1$ and $P_2$ of the initial particles.

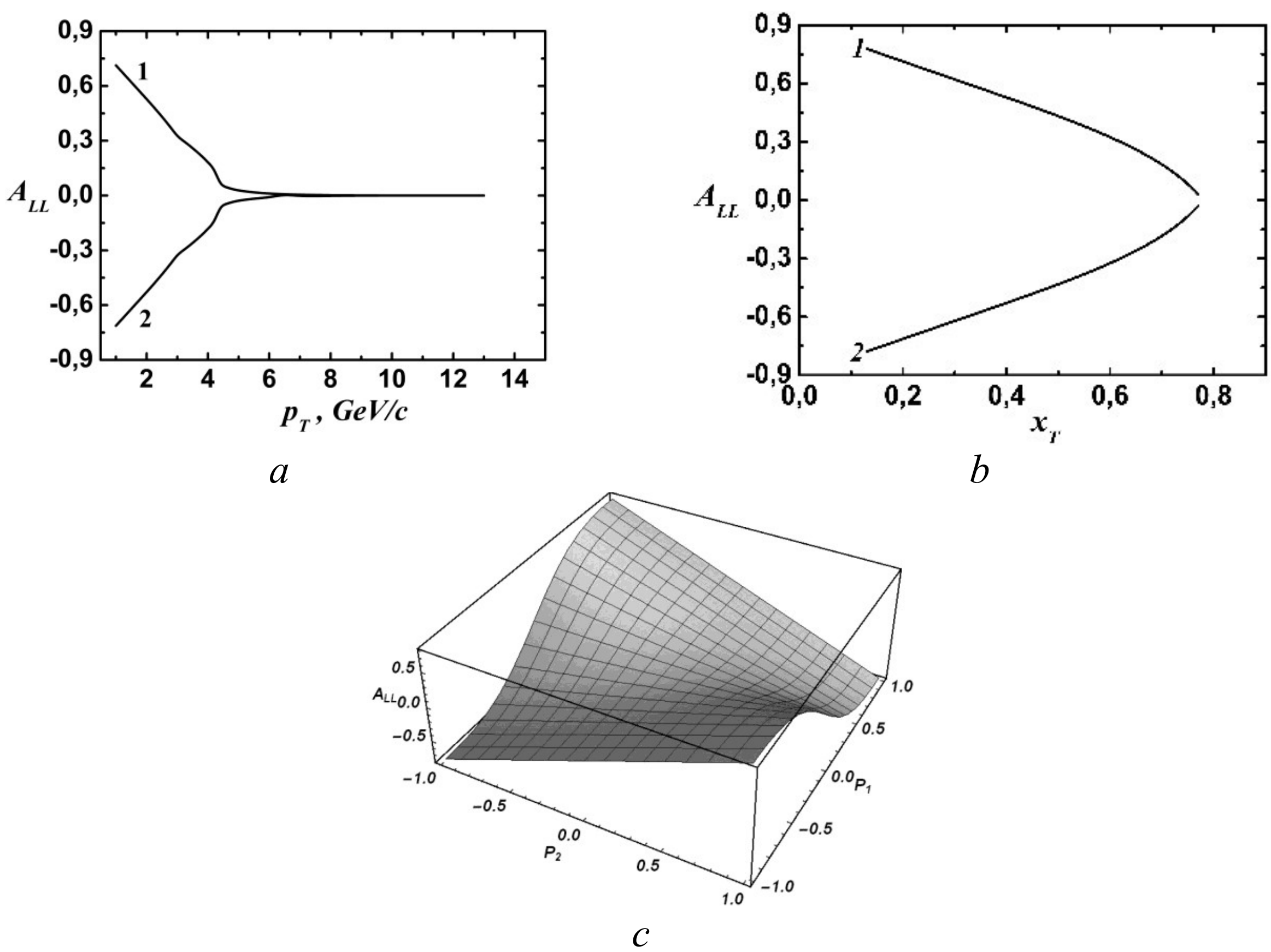

Fig.9($a,b,c$) The dependences of doublespin asymmetry subprocess Compton scattering quark-gluon $qg \rightarrow q\gamma$calculated in NLO at polarization order $P_1$, $P_2$=0.9, -0.9 (curve 1), and at $P_1=P_2=\pm0.9$ (curve 2) on transverse momentum produced prompt photon $p_T$ ($a$), $x_T$ ($b$) and 3-D plot of the dependence of doublespin asymmetry on polarization order $P_1$ and $P_2$ ($c$) at $\sqrt{s}$=27 GeV.

From the dependences (Figure 9(*a,b*)) it is clear that doublespin asymmetry depends differently on the polarizat1ion of the initial particles. The doublespin asymmetry $A_{LL}$ at polarization orders of the initial particles $P_1$, $P_2$=0.9, -0.9 decreases, and at polarization orders $P_1=P_2=\pm0.9$ it decreases with increasing transverse photon momentum $p_T$. This behavior is typical in spin-dependent processes where lower transverse momentum favors higher polarization transfer. At higher $p_T$, gluon contributions dilute the asymmetry, leading to its reduction. At large $p_T$, (correspondingly, at high collision energies $\sqrt{s}$), the value of $A_{LL}$ approaches zero. From the definition of $A_{LL}$, this means that $\sigma^{\uparrow\uparrow}\approx\sigma^{\uparrow\downarrow}$. At high energies, the polarization of the colliding protons has little effect on the differential cross section. A similar dependence is observed in the dependence of doublespin asymmetry on the distribution of transverse momentum $x_T$. For small $x_T$, the asymmetry is large, and as $x_T$ grows, the dilution from gluon interactions suppresses $A_{LL}$. The two curves correspond to different polarization combinations and show a symmetric behavior around zero. As for LO, doublespin asymmetry in NLO has maximum values at opposite polarization $P_1$ and $P_2$ (Figure 9(c)). The asymmetry reaches its maximum when both initial-state polarizations are aligned (same sign) and decreases when they are opposite. The transition from positive to negative asymmetry is continuous, indicating a direct relationship between spin alignment and asymmetry magnitude.

Figure 10(a,b) shows the dependence of the ratio R= $\frac{d\sigma_{NLO_POL}(qg\to q\gamma)}{d\sigma_{NLO}(qg\to q\gamma)}$ on the sum of energy of colliding protons $\sqrt{s}$ and on transverse momentum $p_T$ of photon.

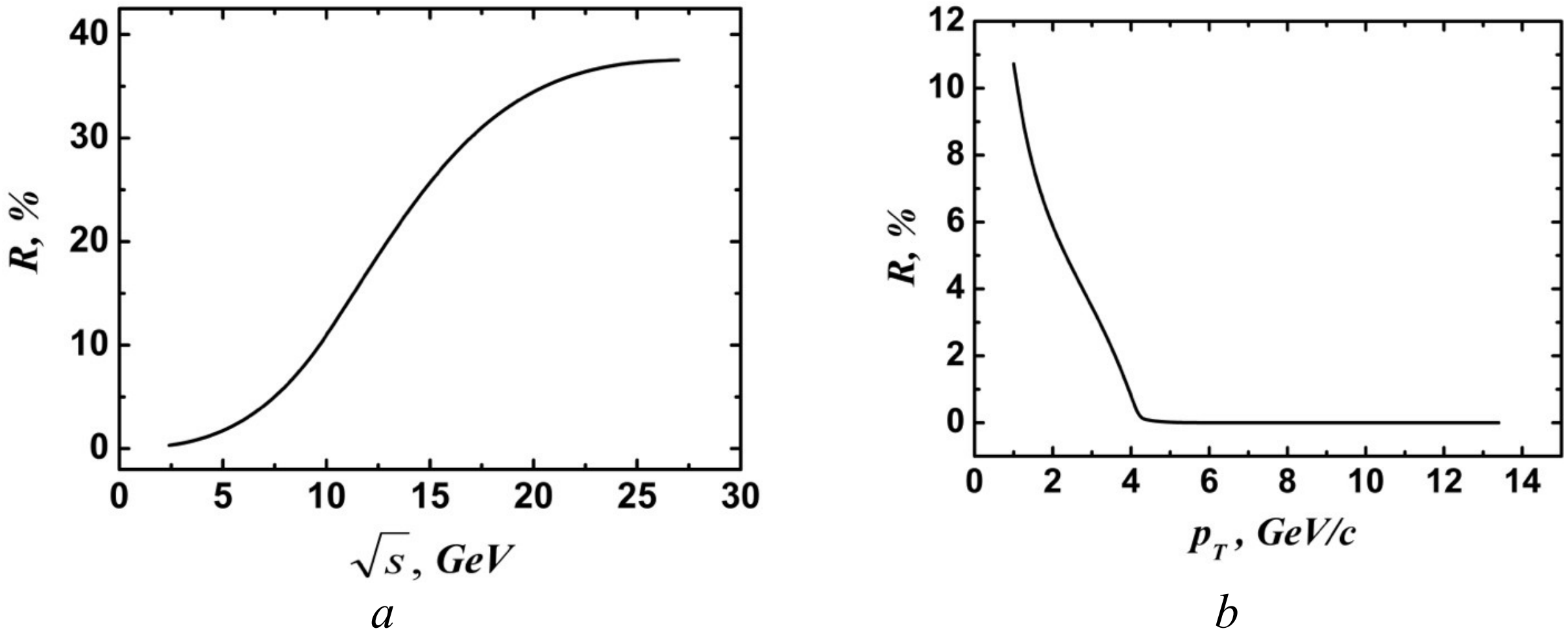

Fig.10(*a*,*b*). Dependence of the ratio $R = \frac{d\sigma_{NLO_POL}(qg\to q\gamma)}{d\sigma_{NLO}(qg\to q\gamma)}$ on sum of energy of colliding protons $\sqrt{s}$ at $P_1P_2$ =-0.81 (curve 1), $P_1P_2$ =0.81 (curve 2) (a) and on the transverse momentum of prompt photons $p_T$ at $\sqrt{s}$=27 GeV.

Depending on the polarization of the colliding particles, polarization can increase or decrease the differential cross-section. The influence of polarization on the differential cross-section of NLO calculations becomes significant at high energies of colliding particles $\sqrt{s}$. Figure 10($a$) shows the dependence of the differential cross-section calculated for the NLO at $P_1P_2$= -0.81, as can be seen, the influence of polarization on the calculations for the NLO becomes significant at high energies of colliding particles $\sqrt{s}$.

As seen from Figure 10($b$) ratio $R = \frac{d\sigma_{NLOPOL}(qg\to q\gamma)}{d\sigma_{NLO}(qg\to q\gamma)}$ decreases with increasing of transverse momentum $p_T$ of prompt photon. The influence of polarization on the differential cross-section of NLO calculations becomes significant at small values of $p_T$.

## CONCLUSIONS

Our calculations showed the importance of NLO calculations for the accurate theory of prompt photon production in Compton scattering of quark-gluon in proton-proton collision.

The dependence of differential cross-section on sum of energy of colliding protons $\sqrt{s}$ calculated in NLO is significant at large values of energy $\sqrt{s}$.

The difference between the differential cross-sections calculated at LO and NLO is small at low and high $p_T$.

The photon has a high probability of scattering along the collision axis of proton. Scattering angles are 16 and 164 degrees.

From the dependences of ratio $R = \frac{d\sigma_{NLO}}{d\sigma_{LO}}$ on the sum of the energies of the colliding initial particles $\sqrt{s}$, the transverse momentum $p_T$ of the prompt photons, the cosine of the scattering angle *Cos(θ)*, the rapidity $y$ of the prompt photons it can be seen that the LO corrections increasingly affect the description of the process. This corresponds to pQCD, in which the higher-order terms give larger radiative corrections at higher energies of the partons.

Comparing the dependences of the differential cross-section calculated in LO and the NLO, it is clear that the polarization of protons has a stronger effect on NLO than on LO. The reason for this may be: firstly, new Feynman diagrams appear in NLO calculations that include corrections for the virtual loop. These new diagrams often produce contributions with opposite signs compared to the LO diagrams, resulting in a decrease in the overall cross-section of the process. However, for specific observable quantities such as polarization asymmetry, this suppression may not be as effective, leading to a stronger relative influence of the remaining polarization-sensitive terms. Second, NLO calculations account for additional kinematic configurations compared to LO, including higher order parton scattering and gluon emission. These configurations may be more sensitive to proton spin orientation, resulting in greater polarization dependence compared to the more constrained kinematic pattern in LO.

Comparison of our calculations with NLO calculations carried out at the LHC and the American Tevatron showed that the NLO is approximately 15% and 35% of LO for the NICA and LHC, American Tevatron energies, correspondingly.

The results obtained indicate the need to take into account NLO contributions when simulation and analyzing experimental data on the process of prompt photon production in proton-proton collisions at NICA and FAIR energies.

## APPENDIX

## FEYNMAN DIAGRAMS AND MATRIX ELEMENTS OF THE PROCESS

The Figure 1(a)-1(k) shows the Feynman diagrams of the process.

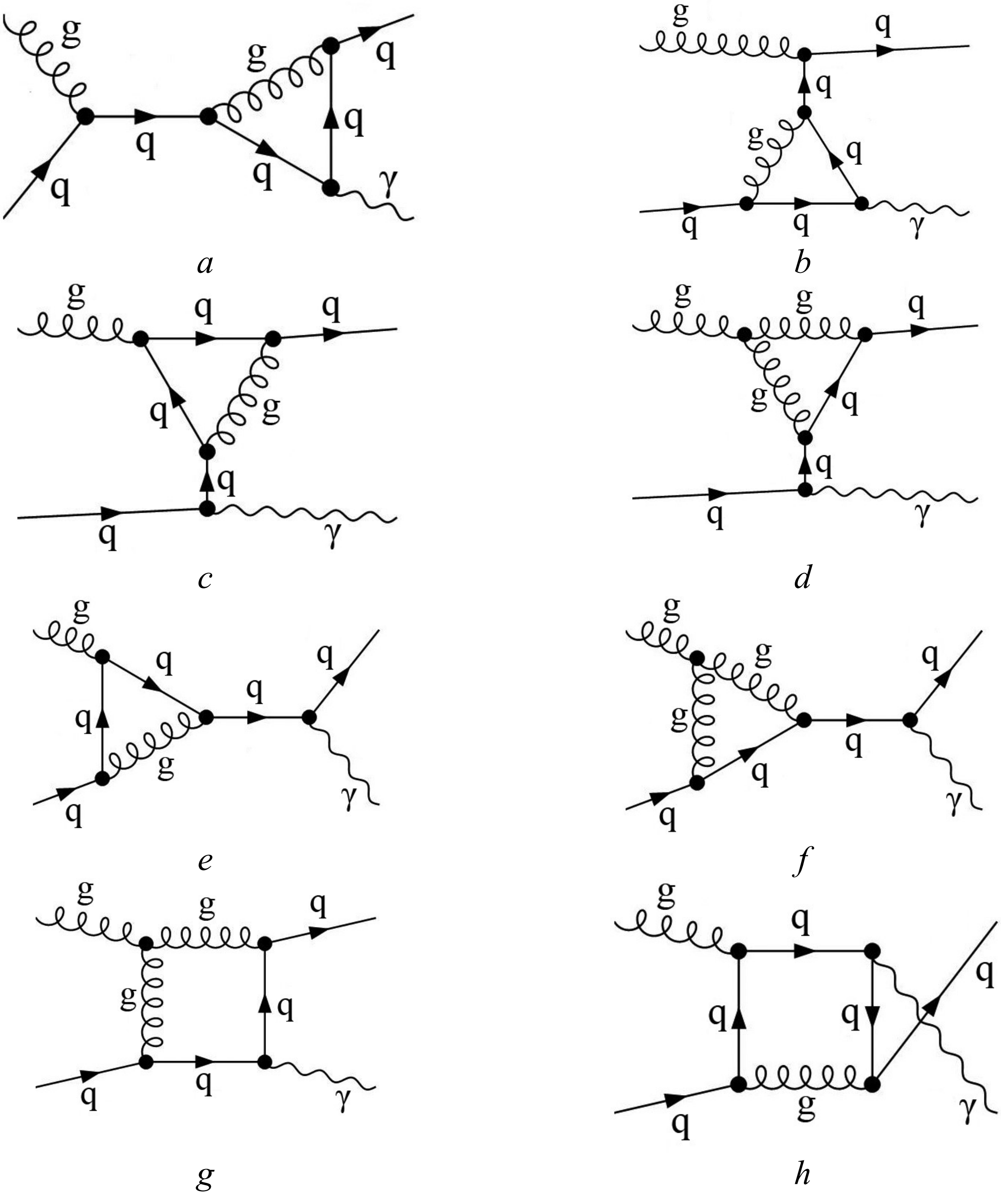

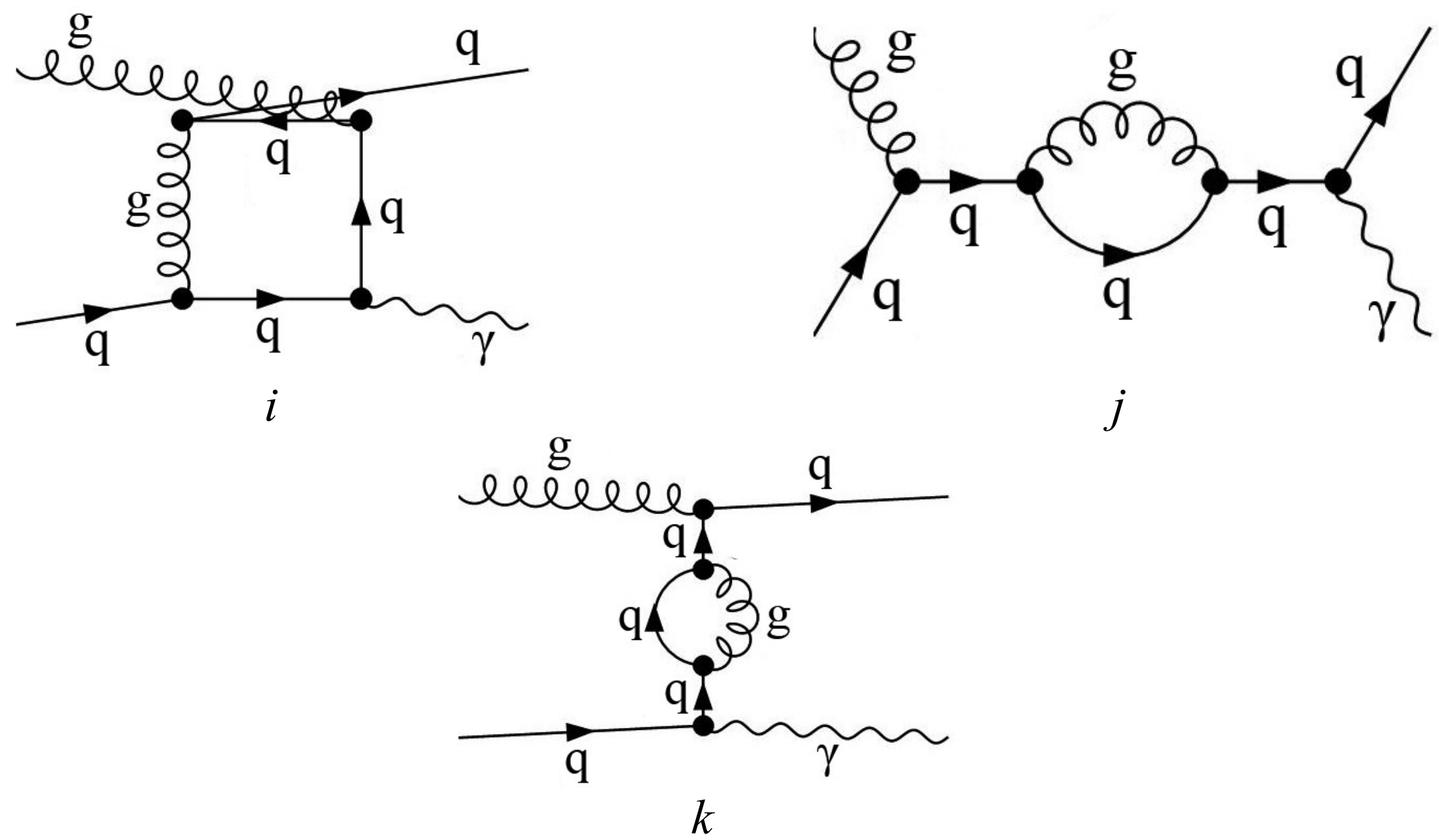

Fig.1(*a,b,c,d,e,f,g,h,i,j,k*) Feynman diagrams of the one-loop corrections to the subprocess of mixed Compton scattering quark-gluon $qg \to q\gamma$

As seen Figure 1(*a,b,c,d,e,f,g,h,i,j,k*) there are the virtual gluon exchange corrections to the lowest order process $qg \to q\gamma$.

The matrix elements of the Feynman diagrams of the $qg \to q\gamma$ subprocess in the next-to-leading order, written using FeynCalc, have the following form:

$$M_1 = \frac{iee_q g_s^3 T^{Glu1}_{Col5Col2} T^{Glu5}_{Col3Col6} T^{Glu5}_{Col6Col5} \bar{u}(p_2)\gamma\hat{q}\hat{\varepsilon}^*(k_1)(\hat{k}_1-\hat{q})\gamma(\hat{k}_1+\hat{p}_2)\hat{\varepsilon}(k_2)u(p_1)}{144\pi^4(k_1+p_2)^2q^2(p_2+q)^2(q-k_1)^2},$$

$$M_2 = \frac{iee_q g_s^3 T^{Glu5}_{Col6Col2} T^{Glu1}_{Col3Col5} T^{Glu5}_{Col5Col6} \bar{u}(p_2)\hat{\varepsilon}(k_2)(\hat{p}_1-\hat{k}_1)\gamma(q-\hat{k}_1)\hat{\varepsilon}^*(k_1)\hat{q}\gamma u(p_1)}{144\pi^4(k_1-p_1)^2q^2(q-p_1)^2(q-k_1)^2},$$

$$M_3 = \frac{iee_q g_s^3 T^{Glu5}_{Col6Col2} T^{Glu5}_{Col3Col5} T^{Glu1}_{Col5Col6} \bar{u}(p_2)\gamma\hat{q}\hat{\varepsilon}(k_2)(\hat{p}_1-\hat{k}_1-\hat{p}_2+\hat{q})\gamma(\hat{p}_1-\hat{k}_1)\hat{\varepsilon}^*(k_1)u(p_1)}{144\pi^4(k_1-p_1)^2q^2(q-p_2)^2(p_1-k_1-p_2+q)^2},$$

$$M_4 = \frac{ee_q g_s^3 T^{Glu6}_{Col5Col2} T^{Glu5}_{Col3Col5} f^{Glu1Glu5Glu6}\big(p_1\varepsilon(k_2) - k_1\varepsilon(k_2) - p_2\varepsilon(k_2) + 2q\varepsilon(k_2)\big)\bar{u}(p_2)\gamma(\hat{p}_2-\hat{q})\gamma(\hat{p}_1-\hat{k}_1)\hat{\varepsilon}^*(k_1)u(p_1)}{144\pi^4(k_1-p_1)^2q^2(q-p_2)^2(p_1-k_1-p_2+q)^2} +$$
$$+ \frac{ee_q g_s^3 T^{Glu6}_{Col5Col2} T^{Glu5}_{Col3Col5} f^{Glu1Glu5Glu6}\bar{u}(p_2)(\hat{k}_1+\hat{k}_2-\hat{p}_1+\hat{p}_2-\hat{q})(\hat{p}_2-\hat{q})\hat{\varepsilon}(k_2)(\hat{p}_1-\hat{k}_1)\hat{\varepsilon}^*(k_1)u(p_1)}{144\pi^4(k_1-p_1)^2q^2(q-p_2)^2(p_1-k_1-p_2+q)^2} +$$
$$+ \frac{ee_q g_s^3 T^{Glu6}_{Col5Col2} T^{Glu5}_{Col3Col5} f^{Glu1Glu5Glu6}\bar{u}(p_2)\hat{\varepsilon}(k_2)(\hat{p}_2-\hat{q})(-\hat{k}_2-\hat{q})(\hat{p}_1-\hat{k}_1)\hat{\varepsilon}^*(k_1)u(p_1)}{144\pi^4(k_1-p_1)^2q^2(q-p_2)^2(p_1-k_1-p_2+q)^2},$$

$$M_5 = \frac{iee_q g_s^3 T^{Glu5}_{Col5Col2} T^{Glu5}_{Col3Col6} T^{Glu1}_{Col6Col5} \bar{u}(p_2)\hat{\varepsilon}^*(k_1)(\hat{k}_1+\hat{p}_2)\gamma(\hat{k}_1-\hat{p}_1+\hat{p}_2-\hat{q})\hat{\varepsilon}(k_2)u(p_1)}{144\pi^4(k_1+p_2)^2q^2(q+p_1)^2(p_1-k_1-p_2+q)^2},$$

$$M_6 = \frac{ee_q g_s^3 T^{Glu5}_{Col5Col2} T^{Glu6}_{Col3Col5} f^{Glu1Glu5Glu6}\big(p_1\varepsilon(k_2) - k_1\varepsilon(k_2) - p_2\varepsilon(k_2) + 2q\varepsilon(k_2)\big)\bar{u}(p_2)\hat{\varepsilon}^*(k_1)(\hat{p}_2+\hat{k}_1)\gamma(\hat{p}_1+\hat{q})\gamma u(p_1)}{144\pi^4(k_1+p_2)^2q^2(q+p_1)^2(p_1-k_1-p_2+q)^2}$$
$$+ + \frac{ee_q g_s^3 T^{Glu5}_{Col5Col2} T^{Glu6}_{Col3Col5} f^{Glu1Glu5Glu6}\bar{u}(p_2)\hat{\varepsilon}^*(k_1)(\hat{k}_1+\hat{p}_2)(-\hat{k}_2-\hat{q})(\hat{p}_1+\hat{q})\hat{\varepsilon}(k_2)u(p_1)}{144\pi^4(k_1+p_2)^2q^2(q+p_1)^2(p_1-k_1-p_2+q)^2} +$$
$$+ \frac{ee_q g_s^3 T^{Glu5}_{Col5Col2} T^{Glu6}_{Col3Col5} f^{Glu1Glu5Glu6}\bar{u}(p_2)\hat{\varepsilon}^*(k_1)(\hat{k}_1+\hat{p}_2)\hat{\varepsilon}(k_2)(\hat{p}_1+\hat{q})(\hat{k}_1+\hat{k}_2-\hat{p}_1+\hat{p}_2-\hat{q})u(p_1)}{144\pi^4(k_1+p_2)^2q^2(q+p_1)^2(p_1-k_1-p_2+q)^2},$$

$$M_7 = \frac{ee_q g_s^3 T_{Col5Col2}^{Glu5} T_{Col3Col5}^{Glu6} f^{Glu1Glu5Glu6}\left(p_1\varepsilon(k_2) - k_1\varepsilon(k_2) - p_2\varepsilon(k_2) + 2q\varepsilon(k_2)\right)\bar{u}(p_2)\gamma\left(\hat{p}_1 - \hat{k}_1 + \hat{q}\right)\hat{\varepsilon}^*(k_1)(\hat{p}_1 + \hat{q})\gamma u(p_1)}{144\pi^4 q^2 (q+p_1)^2 (p_1 - k_1 + q)^2 (p_1 - k_1 - p_2 + q)^2}$$

$$+ + \frac{ee_q g_s^3 T_{Col5Col2}^{Glu5} T_{Col3Col5}^{Glu6} f^{Glu1Glu5Glu6}\bar{u}(p_2)\left(-\hat{k}_2 - \hat{q}\right)\left(\hat{p}_1 - \hat{k}_1 + \hat{q}\right)\hat{\varepsilon}^*(k_1)(\hat{p}_1 + \hat{q})\hat{\varepsilon}(k_2)u(p_1)}{144\pi^4 q^2 (q+p_1)^2 (p_1 - k_1 + q)^2 (p_1 - k_1 - p_2 + q)^2} +$$

$$+ \frac{ee_q g_s^3 T_{Col5Col2}^{Glu5} T_{Col3Col5}^{Glu6} f^{Glu1Glu5Glu6}\bar{u}(p_2)\hat{\varepsilon}(k_2)\left(\hat{p}_1 - \hat{k}_1 + \hat{q}\right)\hat{\varepsilon}^*(k_1)(\hat{p}_1 + \hat{q})\left(\hat{k}_1 + \hat{k}_2 - \hat{p}_1 + \hat{p}_2 - \hat{q}\right)u(p_1)}{144\pi^4 q^2 (q+p_1)^2 (p_1 - k_1 + q)^2 (p_1 - k_1 - p_2 + q)^2},$$

$$M_8 = \frac{-iee_q g_s^3 T_{Col5Col2}^{Glu5} T_{Col3Col6}^{Glu5} T_{Col6Col5}^{Glu1}\bar{u}(p_2)\gamma(\hat{p}_2 - \hat{p}_1 - \hat{q})\hat{\varepsilon}^*(k_1)\left(\hat{k}_1 - \hat{p}_1 + \hat{p}_2 - \hat{q}\right)\hat{\varepsilon}(k_2)\hat{q}\gamma u(p_1)}{144\pi^4 q^2 (q+p_1)^2 (p_1 - p_2 + q)^2 (p_1 - k_1 - p_2 + q)^2},$$

$$M_9 = \frac{iee_q g_s^3 T_{Col6Col2}^{Glu5} T_{Col3Col5}^{Glu5} T_{Col5Col6}^{Glu1}\bar{u}(p_2)\gamma\hat{q}\hat{\varepsilon}(k_2)\left(\hat{p}_1 - \hat{k}_1 - \hat{p}_2 - \hat{q}\right)\hat{\varepsilon}^*(k_1)(\hat{p}_1 - \hat{p}_2 + \hat{q})\gamma\hat{q}\gamma u(p_1)}{144\pi^4 q^2 (q-p_2)^2 (p_1 - p_2 + q)^2 (p_1 - k_1 - p_2 + q)^2},$$

$$M_{10} = \frac{iee_q g_s^3 T_{Col5Col2}^{Glu1} T_{Col3Col6}^{Glu5} T_{Col6Col5}^{Glu5}\bar{u}(p_2)\hat{\varepsilon}^*(k_1)\left(\hat{k}_1 + \hat{p}_2\right)\gamma\hat{q}\gamma\left(\hat{k}_1 + \hat{p}_2\right)\hat{\varepsilon}(k_2)u(p_1)}{144\pi^4 q^2 (k_1+p_2)^4 (q - k_1 - p_2)^2},$$

$$M_{11} = \frac{-iee_q g_s^3 T_{Col6Col2}^{Glu5} T_{Col3Col5}^{Glu1} T_{Col5Col6}^{Glu5}\bar{u}(p_2)\hat{\varepsilon}(k_2)\left(\hat{p}_1 - \hat{k}_1\right)\gamma\hat{q}\gamma\left(\hat{p}_1 - \hat{k}_1\right)\hat{\varepsilon}^*(k_1)u(p_1)}{144\pi^4 q^2 (k_1-p_1)^4 (q - k_1 + p_1)^2}.$$

NLO calculation has been carried out using dimensional regularization and the Minimal Supersymmetric Standard Model (MSSM) scheme [21],

In the parton level the Mandelstam invariants of the subprocess are the following:

$$\hat{s} = (k_1 + p_1)^2, \quad \hat{t} = (k_1 - k_2)^2, \quad \hat{u} = (p_1 - k_2)^2.$$